\documentclass[aps,twocolumn,superscriptaddress,a4paper,floatfix]{revtex4-2}
\usepackage[latin1]{inputenc}
\usepackage{graphicx}
\usepackage{amsmath,amssymb}
\usepackage{hyperref}
\usepackage{color}

\newcommand \beq{\begin{eqnarray}}
\newcommand \eeq{\end{eqnarray}}

\begin{document}

\title{Anomalous  supercurrents in the presence of particle losses  }

\author{Risa Ogino}
\affiliation{Department of Materials Science, Waseda University, Tokyo 169-8555, Japan}
\author{Shun Uchino}
\affiliation{Department of Materials Science, Waseda University, Tokyo 169-8555, Japan}
\affiliation{Department of Electronic and Physical Systems,
Waseda University, Tokyo 169-8555, Japan}


\begin{abstract}
We show that supercurrent properties in a superfluid or superconducting junction are significantly modified by
single-particle losses in a conduction channel. 
In the presence of a spin-independent
particle loss, we find regimes where
the Josephson current $I_N(\phi)$ can develop 
additional zeros within $0<\phi<\pi$ 
and the direction of the supercurrent is reversed. 
Although the region is narrow, we also find a regime
in which  
the critical current is enhanced by dissipation.
Such anomalous behaviors in the Josephson current are attributed to 
a subtle interplay between the contribution 
that is present regardless of dissipation
and the unconventional one that is absent without dissipation
and is associated with the quantum jump term of 
the Gorini-Kossakowski-Sudarshan-Lindblad (GKSL) master equation.
In the presence of a spin-selective particle loss, 
it is shown that a dissipation-induced spin supercurrent and its reversal occur.
The proposed system is analyzed by means of
the Keldysh field theory approach based on
the GKSL master equation and
may be realized in
 ultracold atomic gases and solid-state systems.

\end{abstract}


\maketitle

\section{Introduction}
The Josephson effect that is the flow of a supercurrent  through a weak link without a biased voltage is an outstanding yet ubiquitous phenomenon in physics~\cite{barone1982}. 
In a simple superconducting junction where an insulator is sandwiched  between conventional superconductors, 
the DC current usually obeys
\beq
I=I_C\sin\phi,
\label{eq:josephson}
\eeq 
with critical current $I_C$ and relative phase between two superconductors $\phi$.
The Josephson effect has also been examined in superfluid Helium~\cite{vollhardt2013superfluid}, 
elementary particle physics~\cite{nitta2015josephson},
and ultracold atomic gases~\cite{pethick}. 
In addition to the importance in fundamental physics,
it plays a crucial role as a building block of
quantum technology such as 
superconducting quantum interference device~\cite{fagaly2006superconducting} 
and superconducting qubit~\cite{kjaergaard2020superconducting}.

One of the vital questions in superconducting junctions is
how nonequilibrium disturbance
affects   the Josephson effects~\cite{RevModPhys.76.411}.
The prominent example is the Caldeira-Leggett model
originally introduced to explain the dissipation effect in Josephson junctions.
There, the Ohmic resistance in the junctions is incorporated
as the coupling effect with harmonic oscillators that model an environment~\cite{PhysRevLett.46.211}. 
Another remarkable example, which is relevant to the present study, is the presence of the supercurrent reversal phenomenon
where a dirty superconducting junction is
driven out of equilibrium due to a coupling to a normal current~\cite{baselmans1999}.
The current reversal phenomenon is also known as $\pi$-junction since
the current-phase relation shifts by
$\pi$~\cite{bulaevskii1977,PhysRevB.36.235}.
The $\pi$-junction has originally been realized in a  variety of equilibrium junctions such as ferromagnetic Josephson junction~\cite{PhysRevLett.86.2427},
Josephson junction with unconventional superconductors~\cite{RevModPhys.67.515}, and superconducting quantum dot~\cite{cleuziou2006}, 
and has also attracted attention in terms of quantum computation such as implementation of a qubit with a long coherence time~\cite{PhysRevLett.95.097001}.

\begin{figure}[htbp]
\centering
\includegraphics[width=8.5cm]{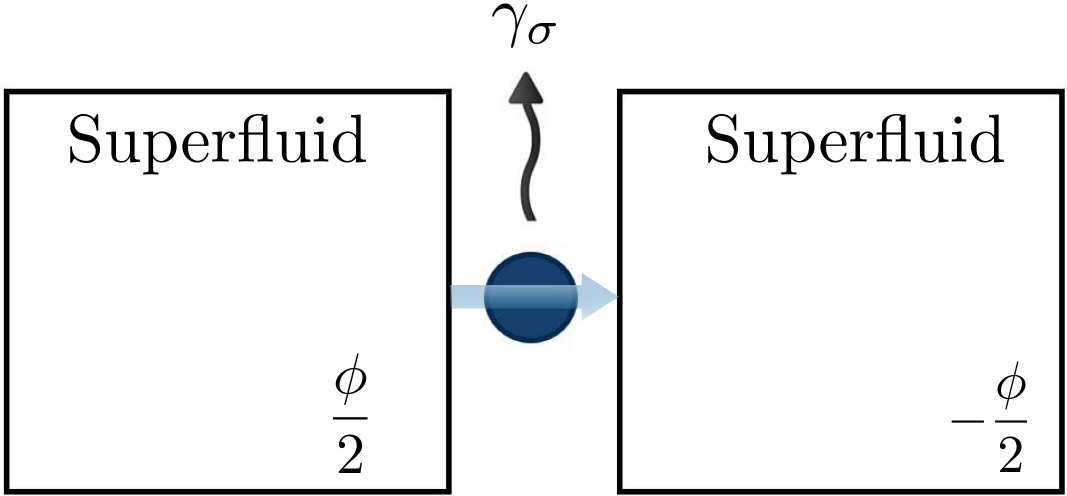}
\caption{\label{fig1} Two-terminal Josephson junction system discussed in this work. A particle current between superfluids or superconductors induced by a relative phase bias $\phi$
passes through a quantum dot 
where single particle losses occur.}
\end{figure}

Nowadays, ultracold atomic gases provide an ideal toolbox to explore 
the Josephson junctions~\cite{krinner2017}.
By controlling an interatomic interaction, one can realize fermionic superfluids
whose transport properties have already been examined in mesoscopic setups such as quantum point contact~\cite{husmann2015} and junction systems~\cite{valtolina2015}.
Josephson current properties in such gases can also be examined through current-biased junctions, which have successfully been
implemented~\cite{levy2007,kwon2020,PhysRevLett.126.055301}.  
Furthermore, a unique capability in ultracold atomic gases allows
for dissipation control in quantum many-body systems~\cite{schafer2020}.
In the context of mesoscopic transport, dissipative transport induced by a single particle loss has been measured~\cite{PhysRevA.100.053605}.
There, a spin-selective dissipation takes place in the mesoscopic channel region in a quantum point contact 
and induces the disappearance of the quantized conductance.
Since the similar experiment under superfluid reservoirs has also been realized with the Feshbach resonance,
which showed that non-linear current-bias curves
arising from the multiple Andreev reflections are weakened
by dissipation~\cite{PhysRevLett.130.200404},
it would be intriguing to examine how such a dissipation affects Josephson current properties.

Another notable development in Josephson junctions is a connection
with non-Hermitian physics~\cite{ashida2020non}.
Recent theoretical works have demonstrated that non-Hermitian effects mimicking dissipation modify the Josephson 
currents~\cite{PhysRevB.109.214514,PhysRevB.110.L201403,PhysRevLett.133.086301,pino2024thermodynamics,10.1063/5.0215522,PhysRevB.110.235426,PhysRevB.111.064517,PhysRevLett.134.156601,sten2025angle}. The non-Hermitian Hamiltonian naturally emerges from the non-unitary evolution term in the Gorini-Kossakowski-Sudarshan-Lindblad(GKSL) master equation~\cite{breuer2002theory}.
At the same time, the full time evolution in the GKSL master equation
also contains the so-called quantum jump term whose inclusion is
required for description of generic Markovian dynamics 
in open quantum systems.
Therefore, it is fundamentally important to look at 
the dissipative Josephson effects at the full GKSL level.
We note that such an analysis has direct relevance to experiments of ultracold atomic gases
whose dissipation effects can usually be captured 
within the GKSL master equation~\cite{daley2014quantum}.

Motivated by recent experimental progress in ultracold atomic gases and
by non-Hermitian treatments of dissipative Josephson junctions,
we investigate  a mesoscopic Josephson junction system shown in Fig.~\ref{fig1}: 
a non-interacting quantum dot accompanied by a single particle loss is coupled  to two superfluid or superconducting reservoirs.
We note that such a system has recently been harnessed to explain
transport properties in the lossy superfluid point contact
realized in ultracold atomic gases~\cite{PhysRevLett.130.200404,PhysRevResearch.5.033095}.
By using the Keldysh formalism that allows to analyze 
the systems governed by the GKSL master equation,
we show that spin-independent loss can generate
additional zero crossings in the particle-current phase relation,
reverse the direction of the supercurrent, and within a narrow parameter window, enhance the critical current.
For spin-selective loss, a spin supercurrent is generated
and can likewise reverse its direction.
We trace these phenomena to unconventional current contributions
that vanish in the absence of loss.
For the particle current, the unconventional contribution competes
with the conventional Josephson contribution, whereas the spin supercurrent
consists entirely of an unconventional contribution.
These unconventional terms arise from the dissipative Keldysh structure
of the full GKSL dynamics and include a contribution associated with
the quantum-jump term, which is absent from an effective non-Hermitian treatment.

The structure of the paper is as follows.
Section~II introduces the model and the Keldysh formalism of the GKSL dynamics. 
By applying this formalism to 
the superconducting junction, we obtain the Josephson current formulae
that allow us to discuss the dissipation effects.
Section~III presents the current-reversal mechanism, its parameter regime,
and loss-induced spin supercurrent.
Section~IV is devoted to discussion and outlook.
Appendices A-D give
the Green's functions,
Dyson equation in the Keldysh formalism, current expressions,
and additional results for spin-selective loss.

\section{Formulation of the problem}

To uncover prototypical roles of particle losses 
in Josephson junctions, we 
consider the system in which 
two superfluid or superconducting reservoirs
are connected via a non-interacting quantum dot~(Fig.~\ref{fig1}).
In the absence of dissipation, the evolution of the density matrix $\rho$ is described by the von Neumann equation  $\partial_{\tau}\rho=-i[H,\rho]$ with the Hamiltonian
\beq
H=H_L+H_R+H_{T}+H_{d}.
\eeq
We model the two reservoirs by the mean-field BCS Hamiltonians
\beq
H_{L(R)}=\sum_{\mathbf{k},\sigma}\xi_{\mathbf{k}}c^{\dagger}_{\mathbf{k}\sigma L(R)}c_{\mathbf{k}\sigma L(R)}\nonumber\\
-\sum_{\mathbf{k}}(\Delta e^{i\phi_{L(R)}}c^{\dagger}_{\mathbf{k}\uparrow L(R)}
c^{\dagger}_{-\mathbf{k}\downarrow L(R)}+h.c.)
\eeq
Here $c_{\mathbf{k}\sigma L(R)}$ annihilates a spin-$\sigma$
fermion in reservoir $L(R)$, 
$\xi_{\mathbf{k}}$ is the single-particle dispersion measured from the common chemical potential,
$\Delta$ is superconducting gap, and $\phi_{L/R}=\pm\frac{\phi}{2}$ is the superconducting phase.
In addition, the tunneling and dot Hamiltonians are
\beq
&&H_{T}=\sum_{\sigma}t[\psi^{\dagger}_{L,\sigma}d_{\sigma}+\psi^{\dagger}_{R,\sigma}d_{\sigma}+\text{h.c.}],\\
&&H_{d}=\sum_{\sigma}\epsilon d^{\dagger}_{\sigma}d_{\sigma}.
\eeq
The local reservoir field $\psi_{L(R),\sigma}$ is evaluated at the contact position,
$d_{\sigma}$ annihilates a dot fermion,
$t$ is the tunneling amplitude between reservoirs and quantum dot, and
$\epsilon$ the onsite energy in the dot~\footnote{We assume that tunneling occurs between the position $\mathbf{r}=\mathbf{0}$
in each reservoir and the quantum dot. Therefore, $\psi_{L(R)\sigma}$ is the shorthand
notation of $\psi_{L(R)\sigma}(\mathbf{0})$.}.
Notice that we adopt units of $\hbar=k_B=1$.
In the presence of a single particle loss at the dot, the system no longer experiences the unitary evolution and instead is described by the following quantum master equation~\cite{breuer2002theory}:
\beq
\partial_{\tau}\rho=-i[H,\rho]+\sum_{\sigma}\gamma_{\sigma}\Big(2d_{\sigma}\rho d^{\dagger}_{\sigma}-\{d^{\dagger}_{\sigma}d_{\sigma},\rho\}\Big),
\label{eq:master}
\eeq
where $\gamma_{\sigma}$ represents  the particle loss rate.

Based on this setup, we derive the current expressions used in this work.
To this end, we first consider the time derivative of
the spin-$\sigma$ particle numbers in each reservoir and in the quantum dot, which are given by
\beq
\dot{N}_{\sigma,L/R}=\text{Tr}[N_{\sigma,L/R}\partial_{\tau}\rho]
&=&-it\langle \psi^{\dagger}_{\sigma,L/R}d_{\sigma}\rangle+h.c.
\eeq
\beq
\dot{N}_{\sigma,d}=it\sum_{j=L,R}\langle \psi^{\dagger}_{\sigma,j}d_{\sigma}\rangle+h.c.-2\gamma_{\sigma}\langle
d^{\dagger}_{\sigma}d_{\sigma}\rangle.
\eeq
By adopting the convention that a current is positive when particles flow from left to right reservoirs, 
the particle and spin currents between reservoirs $I_N$ and $I_S$ 
are defined as follows~\cite{PhysRevA.106.053320,PhysRevLett.130.200404,PhysRevResearch.5.033095}:
\beq
I_{N}&=&\frac{-\dot{N}_{L}+\dot{N}_R}{2}
=\frac{1}{2}\sum_{\sigma}(-\dot{N}_{\sigma,L}+\dot{N}_{\sigma,R})
\nonumber\\
&=&\sum_{\sigma}\text{Re}\Big[it\{\langle \psi^{\dagger}_{L,\sigma}d_{\sigma}\rangle - \langle \psi^{\dagger}_{R,\sigma}d_{\sigma}\rangle\}\Big],
\label{eq:current}
\eeq
\beq
I_{S}&=&\frac{1}{2}
(-\dot{N}_{\uparrow,L}+\dot{N}_{\downarrow,L}
+\dot{N}_{\uparrow,R}-\dot{N}_{\downarrow,R})
\nonumber\\
&=&\text{Re}\Big[it\{\langle \psi^{\dagger}_{L,\uparrow}d_{\uparrow}\rangle - \langle \psi^{\dagger}_{R,\uparrow}d_{\uparrow}\rangle\}
-it\{\langle \psi^{\dagger}_{L,\downarrow}d_{\downarrow}\rangle - \langle \psi^{\dagger}_{R,\downarrow}d_{\downarrow}\rangle\}
\Big].\nonumber\\
\label{eq:scurrent}
\eeq
We point out that although the expressions above are formally equivalent to those without dissipation, expectation values are taken with the density matrix satisfying Eq.~\eqref{eq:master} and therefore fully contain the effects of dissipation.

In the presence of particle losses, particles leak out of the system. The corresponding flow is the loss current defined as
\beq
-\dot{N}&=&-(\dot{N}_L+\dot{N}_R+\dot{N}_d)\nonumber\\
&=&2\sum_{\sigma}\gamma_{\sigma}\langle d^{\dagger}_{\sigma}d_{\sigma
}\rangle
\label{eq:lcurrent}
\eeq
This expression vanishes in the absence of dissipation,
as it should be.

\begin{figure*}[htbp]
    \centering
    \includegraphics[width=2\columnwidth,clip]{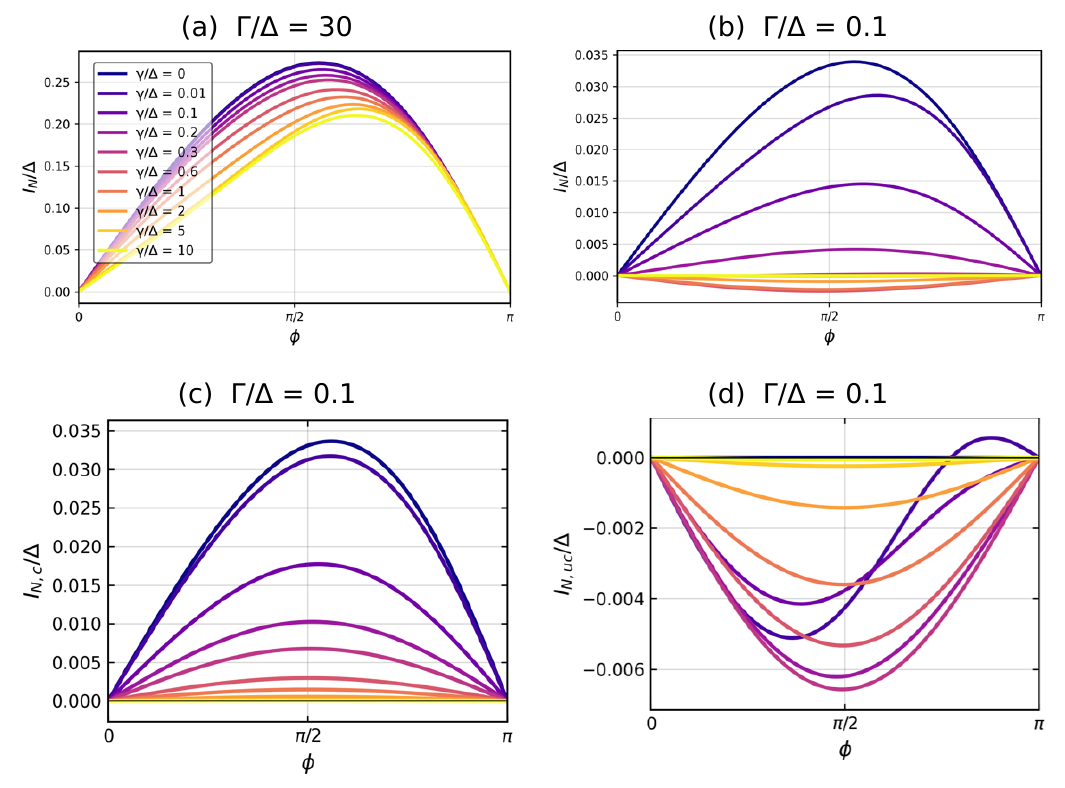}
    \caption{\label{fig2}
    Loss-induced reversal of the particle supercurrent for spin-independent loss, $\gamma_{\uparrow}=\gamma_{\downarrow}$.
    Results are for zero temperature and transmission
    ${\cal T}=0.5$. (a,b) Total particle current $I_N$ for $\Gamma/\Delta=30$ and 0.1, respectively. (c,d) Conventional and unconventional contributions, $I_{N,c}$ and  $I_{N,uc}$, respectively, for $\Gamma/\Delta=0.1$. The curves correspond to the values of $\gamma/\Delta$  listed in panel (a). 
    }
\end{figure*} 

In what follows,
we focus on the behaviors of Eqs.~\eqref{eq:current},~\eqref{eq:scurrent},~\eqref{eq:lcurrent}  in the presence of a relative phase bias between fermionic superconducting or superfluid  reservoirs.  The particle current then corresponds to
the Josephson current, or supercurrent, while the spin current corresponds to the spin supercurrent.

To explore current behaviors,  we adopt the Keldysh formalism, which allows to analyze observables in nonequilibrium~\cite{kamenev2011} 
and open quantum systems~\cite{sieberer2016}.
In addition, we treat the Fermi superfluid in each reservoir within the mean-field theory, which has been used to compare with cold-atom experiments in Ref.~\cite{husmann2015,PhysRevLett.130.200404}.

To relate the average currents with quantities calculable from the Keldysh formalism, we define
the Keldysh components of Green's functions between reservoirs and quantum dot  (see also Appendix A)
\beq
&&\hat{G}^K_{dL(R)}(\tau, \tau') = -i \langle [ \hat{d}(\tau),
\hat{\psi}_{L(R)}^\dagger(\tau') ] \rangle \label{eq:KeldyshdLR},\\
&&\hat{G}^K_{L(R)d}(\tau, \tau') = -i \langle [ \hat{\psi}_{L(R)}( \tau), \hat{d}^\dagger(\tau') ] \rangle \label{eq:KeldyshLRd}.
\eeq
Here we introduce the following Nambu spinors of the fermionic fields~\cite{PhysRevB.54.7366}:
\beq
\hat{\psi}_{L(R)}=\begin{pmatrix}
\psi_{\uparrow,L(R)}\\
\psi^{\dagger}_{\downarrow,L(R)}
\end{pmatrix}, \ \ 
\hat{d}=\begin{pmatrix}
d_{\uparrow}\\
d^{\dagger}_{\downarrow}
\end{pmatrix}, 
\eeq
which provides useful expressions in fermionic superfluids
that exist nonzero anomalous averages such as $\langle\psi_{\uparrow}\psi_{\downarrow}\rangle$.
By using above, $I_N$ and $I_S$ are rewritten as
\begin{align}
I_N =&\frac{1}{4}\mathrm{Tr}\bigg[\hat{\sigma}_z \hat{t}^\dagger \hat{G}^K_{dL}(\tau, \tau) - \hat{\sigma}_z \hat{t} \hat{G}^K_{Ld}(\tau, \tau)\nonumber\\
&-\hat{\sigma}_z \hat{t} \hat{G}^K_{dR}(\tau, \tau)  + \hat{\sigma}_z \hat{t}^\dagger \hat{G}^K_{Rd}(\tau, \tau) \bigg],\\
I_S=&\frac{1}{4}\mathrm{Tr}\bigg[\hat{t}^\dagger \hat{G}^K_{dL}(\tau, \tau)-\hat{t} \hat{G}^K_{Ld}(\tau, \tau)\nonumber\\
&-\hat{t} \hat{G}^K_{dR}(\tau, \tau)+\hat{t}^\dagger \hat{G}^K_{Rd}(\tau, \tau) \bigg],
\end{align}
where
\begin{align}
\hat{t} = 
 \begin{pmatrix}
    t e^{-i\phi/4} & 0 \\
    0 & -t e^{i\phi/4}
\end{pmatrix},
\end{align}
\begin{align}
\hat{\sigma}_z = 
\begin{pmatrix}
    1 & 0 \\
    0 & -1
\end{pmatrix},
\end{align}
and the trace is taken for the Nambu space.
For the purpose of convenience, we performed a gauge transformation that renders the pair potentials real
and transfers the phase bias to the tunneling matrix above~\cite{PhysRevB.54.7366}.
It follows that the above currents are 
determined once Eqs~\eqref{eq:KeldyshdLR} and~\eqref{eq:KeldyshLRd} 
are calculated.

As shown in Appendices B and C,
Eqs~\eqref{eq:KeldyshdLR} and~\eqref{eq:KeldyshLRd} 
can systematically be analyzed with the Dyson equation
in the Keldysh formalism.
Such an analysis reveals that
$I_N$ is given by
\beq
I_N=I_{N,c}+I_{N,uc}.
\label{eq:particle-current}
\eeq
Here $I_{N,c}$ is the contribution that exists even in the absence of dissipation, and $I_{N,uc}$ is the unconventional contribution that only emerges in the presence of dissipation.
Furthermore, the similar analysis for $I_S$
leads to
\beq
I_S=I_{S,uc},
\label{eq:spin-current}
\eeq
where $I_{S,uc}$ represents 
the unconventional contribution that emerges in the presence of spin-selective dissipation.
The absence of the conventional contribution in the spin current
is attributed to the fact that the spin current is not usually generated
by the phase bias yet is allowed owing to the spin-dependent 
particle loss.

We point out that the loss current can also be evaluated from the Keldysh formalism.
By using Green's functions defined in Appendix A,
it can be expressed in terms of the quantum-dot Green's functions as follows:
\beq
-\dot{N}=&i\text{Tr}\Big[\hat{\gamma}\{\hat{G}^R_d(\tau,\tau)
-\hat{G}^A_d(\tau,\tau)\}-\hat{\gamma}\hat{\sigma}_z\hat{G}^K_{d}(\tau,\tau)
\Big],\nonumber\\
\label{eq:lcurrent2}
\eeq
where 
\beq
\hat{\gamma}=\begin{pmatrix}
\gamma_{\uparrow} & 0\\
0 &\gamma_{\downarrow}
\end{pmatrix}
\eeq
This result must be reasonable, because
the loss current is proportional to the distribution of 
the quantum dot.

\section{Dissipation-induced anomalous supercurrents}
We organize our numerical results around three principal effects
of particle loss: reversal of the particle supercurrent,
a modest enhancement of the critical current within a narrow parameter regime, and generation and 
reversal of a spin supercurrent.
All calculations are performed at zero temperature,
so that thermal suppression of the supercurrent is absent.
\par
We define the resonance width
 $\Gamma=2t^2/W$, where $W=1/(\pi\rho)$ and $\rho$ is the normal-state density of states at the Fermi energy. 
The normal-state transmission probability through the channel at the Fermi 
energy is ~\cite{martin2011}
\beq
{\cal T}=\frac{1}{1+(\epsilon/\Gamma)^2}
\eeq
For $\Gamma/\Delta\gg1$, the current is dominated by Andreev
bound states, whereas for $\Gamma/\Delta\lesssim1$ the continuum
contribution becomes important~\cite{martin2011}.
\par
To evaluate the current expressions obtained in Appendix C, we perform
the frequency integrals with a finite cutoff. 
In the numerical calculations, we choose a sufficiently large cutoff such that convergent results are 
obtained.
\par
Because the quantum dot is coupled symmetrically to the two reservoirs, interchanging $L$
and $R$ maps $\phi$ to $-\phi$ and reverses the inter-reservoir currents: $I_{N,S}(-\phi)=-I_{N,S}(\phi)$. Together with $2\pi$ periodicity, this
gives $I_{N,S}(2\pi-\phi)=-I_{N,S}(\phi)$.
By contrast, the loss current is even under phase reversal, 
$-\dot{N}(-\phi)=-\dot{N}(\phi)$.
It is therefore sufficient to present the current-phase
relations over the interval $0\le\phi\le\pi$.

\subsection{GKSL mechanism of the current reversal}
To demonstrate how single-particle loss can give rise to the current reversal
phenomenon, we first consider spin-independent loss, $\gamma_{\uparrow}=\gamma_{\downarrow}\equiv \gamma$.
\par
In the broad-resonance regime $\Gamma/\Delta\gg1$,
increasing $\gamma$ monotonically suppresses the supercurrent without
changing its direction, as shown in Fig.~\ref{fig2}~(a).
This behavior is analogous to the washing out  of coherent Andreev transport
by increasing temperature.
\par
A qualitatively different regime emerges for
$\Gamma/\Delta\lesssim1$.
Although the overall current amplitude tends to decrease
with increasing dissipation, as in the  
$\Gamma/\Delta\gg1$ case,
$I_N$ can take negative values within $0<\phi<\pi$ at intermediate
loss (Fig.~\ref{fig2}~(b)).
The symmetry relation given above then implies corresponding
positive values  for $\pi<\phi<2\pi$.
This sign reversal is characteristic of a loss-induced $\pi$ junction.
\par
The mechanism of the reversal is transparent from the decomposition
in Eq.~\eqref{eq:particle-current}.
The conventional contribution $I_{N,c}$ decreases 
monotonically with loss (Fig.~\ref{fig2}~(c)).
By contrast, $I_{N,uc}$ vanishes at $\gamma=0$, is nonmonotonic in $\gamma$,
and has the opposite sign to $I_{N,c}$ (Fig.~\ref{fig2}~(d)).
Once $|I_{N,uc}|$ exceeds $|I_{N,c}|$, the total current develops
additional zero crossings and reverses its direction.
The unconventional term originates from the dissipative Keldysh structure
of the full GKSL evolution and, in particular, contains the quantum-jump contribution that is absent from an effective non-Hermitian description.

\subsection{Parameter regime and dissipation-enhanced critical current}
\begin{figure}[htbp]
\centering
\includegraphics[width=8.5cm]{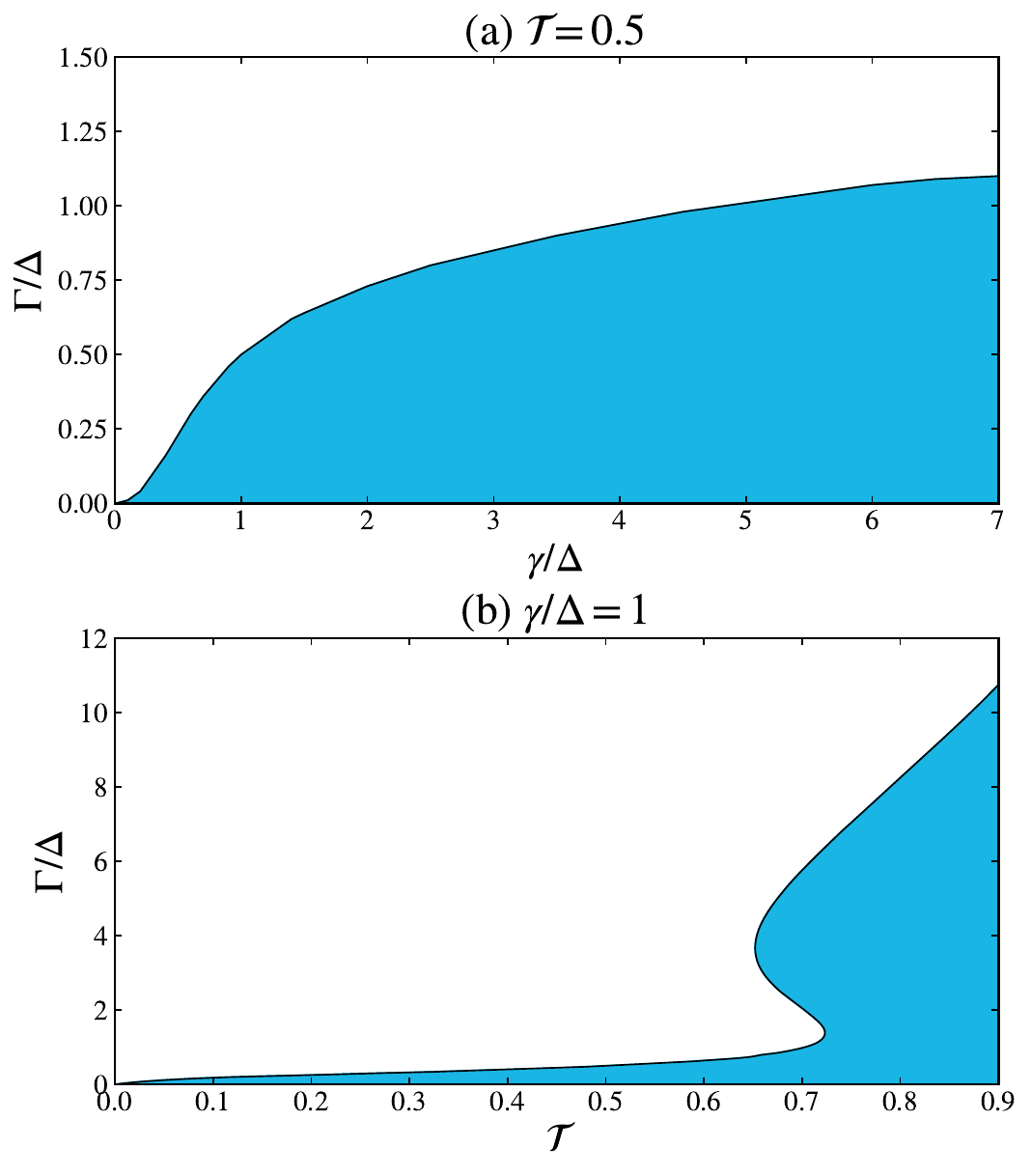}
\caption{\label{fig3}
Phase diagram for anomalous particle-current phase relations under
spin-independent loss.
(a) Fixed transmission $\mathcal{T}=0.5$.
(b) Fixed loss rate $\gamma/\Delta=1$. Blue regions denote
parameters for which $I_N$ exhibits additional zero crossings and current reversal.}
\end{figure}
The phase diagrams in Fig.~\ref{fig3} delineate the parameter
regions in which
the current-phase relation exhibits additional zero crossings
and current reversal.
At fixed transmission ${\cal T}=0.5$, the anomalous behavior is
largely confined to $\Gamma/\Delta\lesssim1$ and 
the anomalous region extends toward larger $\Gamma/\Delta$
 as the loss is increased over the range shown.
At fixed $\gamma/\Delta=1$, high transmission favors the anomalous current-phase relation, with a re-entrant boundary near ${\cal T}\simeq0.7$,
which has also been found in out-of-equilibrium Josephson junction~\cite{baselmans1999}.
These trends are consistent with the mechanism described above:
the broad-resonance current is dominated by $I_{N,c}$,
whereas the unconventional contribution can compete effectively in the 
narrow-resonance, high-transmission regime. 
\par
\begin{figure}[htbp]
\centering
\includegraphics[width=8cm]{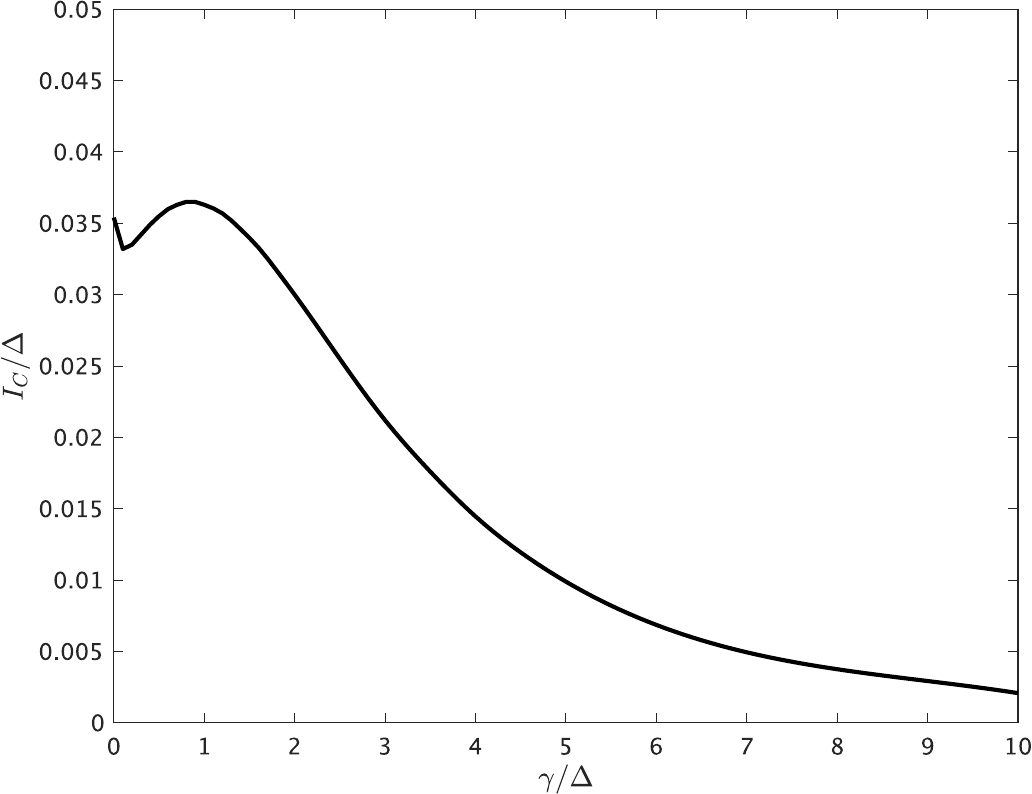}
\caption{\label{fig4} 
Critical particle current $I_C=\max{0\le\phi\le\pi}|I_N(\phi)|$  as a function of
$\gamma/\Delta$ for   $\Gamma/\Delta=1$ and $\mathcal{T}=0.1$.}
\end{figure}

The results above show that the dissipation does not simply
suppress the coherent supercurrent.
The same competition can produce a modest enhancement of the critical current,
which is defined as
\beq
I_C=\max_{0\le\phi\le\pi}|I_N(\phi)|.
\eeq
For $\Gamma/\Delta=1$ and ${\cal T}=0.1$, we find the regime that $I_C$
rather increases with loss and reaches a maximum near $\gamma/\Delta\simeq1$
before decaying at stronger loss (Fig.~\ref{fig4}).
The enhancement is not generic: it occurs only in a narrow
region where the nonmonotonic unconventional contribution outweighs
the simultaneous reduction of the conventional current.
Presenting this result together with the phase diagram makes clear
that it is a secondary consequence of the same competition rather than
an independent mechanism.

\begin{figure*}[htbp]
    \centering
    \includegraphics[width=2\columnwidth,clip]{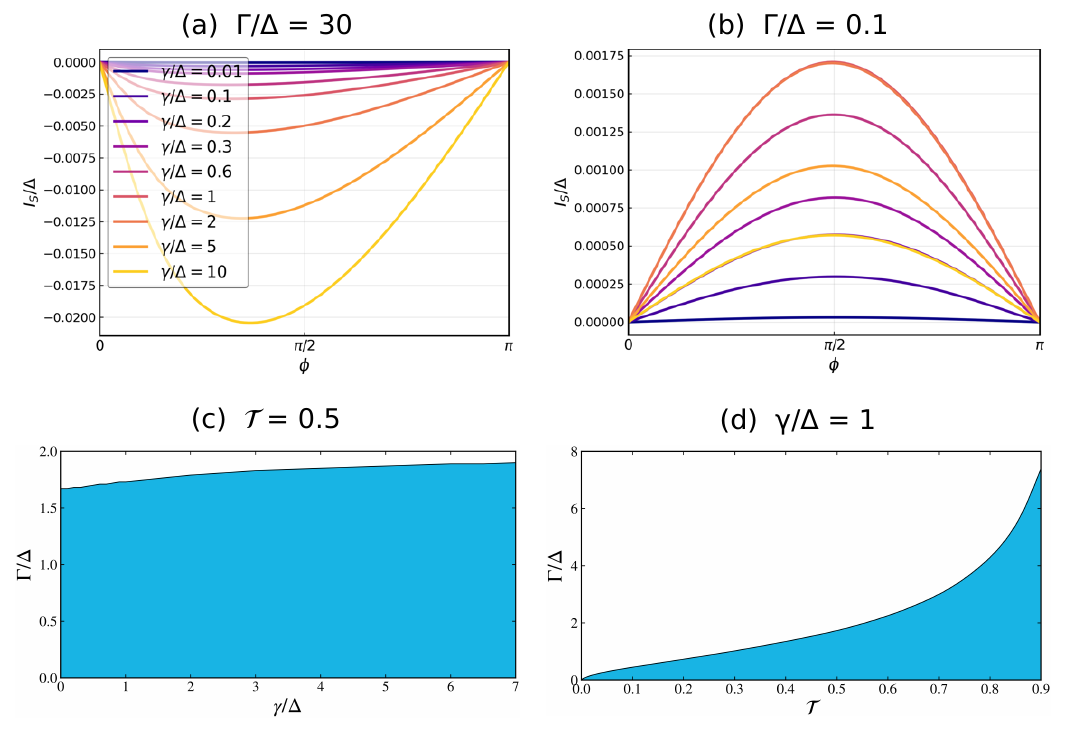}
    \caption{\label{fig5}
    Spin supercurrent generated by spin-selective loss, $\gamma_{\uparrow}=\gamma$ and $\gamma_{\downarrow}=0$.
    (a,b) $I_S(\phi)$ at ${\cal T}=0.5$ for $\Gamma/\Delta=30$
    and 0.1; the curves use the $\gamma/\Delta$ values listed in (a). (c,d) Phase diagrams at fixed ${\cal T}=0.5$
    and fixed $\gamma/\Delta=1$, respectively.
     Blue regions denote the reversed spin-current phase relation. 
    }
\end{figure*} 

\subsection{Spin supercurrent generated by spin-selective loss}
We next move on to the spin-dependent dissipation.  
For the sake of simplicity, we analyze the situation of $\gamma\equiv\gamma_{\uparrow}$ and $\gamma_{\downarrow}=0$. 
\par
The particle current retains the same qualitative distinction
as in the spin-independent case: it is monotonically suppressed for
$\Gamma/\Delta\gg1$ and can become anomalous for $\Gamma/\Delta\lesssim1$.
Because the corresponding supercurrent scans repeat the mechanism
established above, they are provided in Appendix D.
Here we focus on the genuinely new observable, the spin supercurrent.
\par
Equation~\eqref{eq:spin-current} shows that
the spin current consists entirely of the unconventional
contribution and is generated only in the presence of spin-selective dissipation.
In the broad-resonance regime, the induced spin current is 
negative for $0<\phi<\pi$ and positive for $\pi<\phi<2\pi$~(Fig.~\ref{fig5}~(a)).
Because the loss acts directly only on the up-spin component,
it creates an imbalance between the two spin-resolved supercurrents.
For $\Gamma/\Delta\lesssim1$, 
the spin-current phase relation has the sign opposite to that in the broad-resonance
regime over the range of loss shown (Fig.~\ref{fig5}~(b)).
Stronger loss eventually suppresses the generated spin supercurrent
as phase-coherent transport is washed out.
\par
Figures~\ref{fig5}~(c) and (d) delineate the parameter region
in which the spin-current phase relation is reversed.
At fixed ${\cal T}=0.5$, the phase boundary changes only weakly with
loss over the plotted interval, whereas at fixed $\gamma/\Delta=1$
the anomalous region expands rapidly with transmission.
Thus, particle-current reversal and spin-current reversal arise
from the same loss-generated Keldysh contribution.
Because the spin current has no conventional background, 
it provides a particularly direct signature of spin selective dissipation.

\section{Discussion and outlook}
We have shown that local particle loss can qualitatively modify, rather than merely suppress coherent Josephson transport.
Within the full GKSL description, particle loss generates
 an unconventional  contribution to the inter-reservoir current. 
Under spin-independent loss, competition between this contribution and the conventional Josephson current 
produces additional zero crossings, a loss-induced 0-to-$\pi$ transition, and, in a narrow
region, an enhanced critical current. Under spin-selective loss, the 
corresponding unconventional
contribution generates a spin supercurrent and can reverse its direction. 
Thus, the anomalous particle- and spin-current responses originate from
the same dissipative Keldysh structure of the GKSL dynamics.
\begin{figure}[htbp]
\centering
\includegraphics[width=9cm]{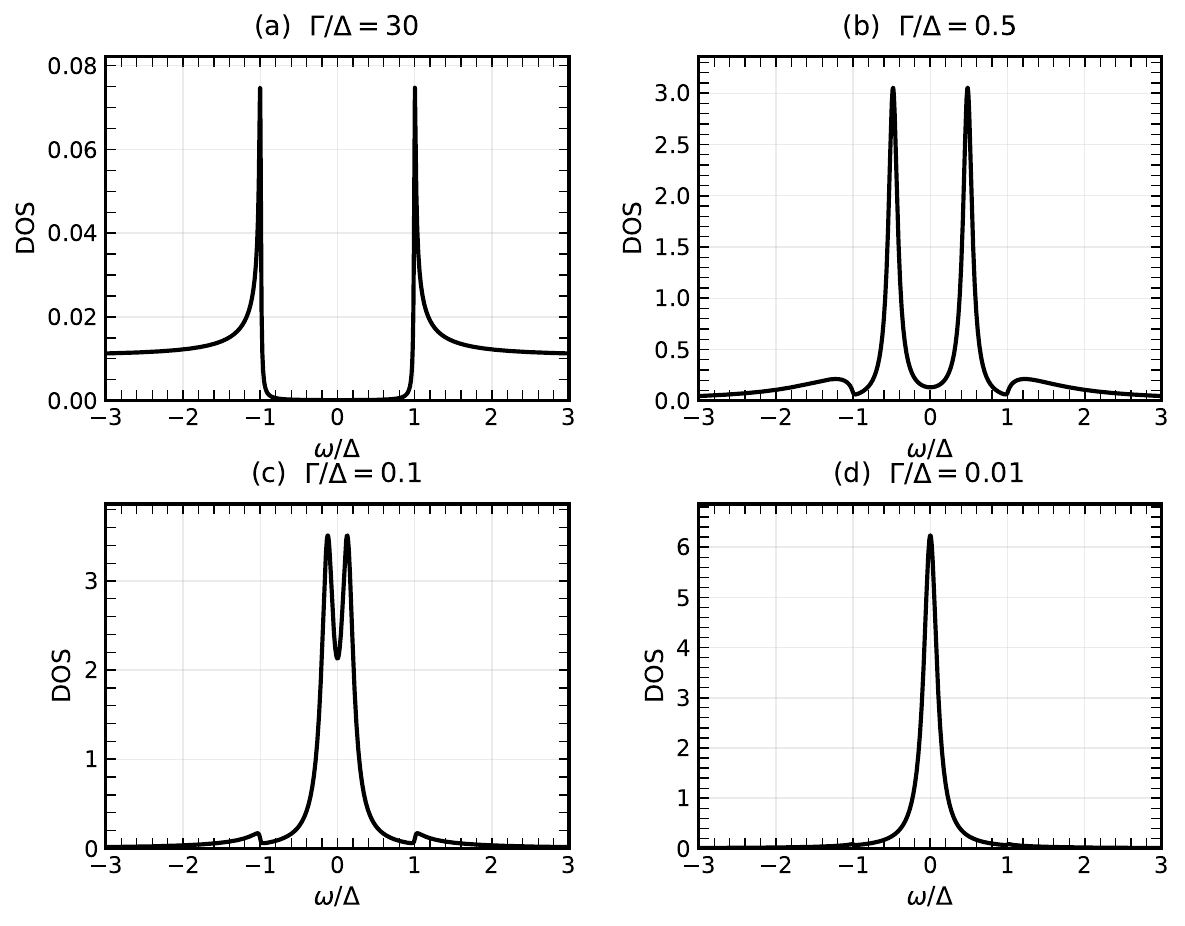}
\caption{\label{fig6} 
Quantum-dot density of states 
in units of $\Delta$ for spin-independent loss at $\gamma/\Delta=0.1$
and $\phi=0$.
Panels (a)-(d) correspond to $\Gamma/\Delta=30, 0.5, 0.1$, and $0.01$,
respectively. }
\end{figure}

Although the current reversal originates from the Keldysh sector of the
GKSL dynamics and therefore cannot be interred from the retarded spectrum alone, the quantum-dot spectral density provides a useful characterization of
the parameter regime in which the anomalous current-phase relation appears.
We define the density of states (DOS) in the quantum, which
is known to contain useful information in superconducting junction~\cite{martin2011}.
\par
Figure~\ref{fig6} shows the DOS defined by
$-\text{Im}[\text{Tr}[\hat{G}^R_d(\omega)]]$.
For a sufficiently large  $\Gamma/\Delta$ (Fig.~\ref{fig6}~(a)),
spectral peaks originating from the Andreev bound states
responsible for the conventional Josephson current 
appear near $\omega=\pm\Delta$ for $\phi=0$.
As $\Gamma/\Delta$ decreases, however, 
the loss-induced spectral peak inside the
superconducting gap becomes non-negligible and
overlaps with the Andreev-bound-state peaks.
Upon further decreasing $\Gamma/\Delta$,
the Andreev-bound-state peaks disappear and
a single peak originating from particle loss in the quantum dot remains.
We  find that the anomalous supercurrent behavior  qualitatively emerges when the Andreev-bound-state peaks are washed out and the loss-induced spectral peak becomes dominant. 
This spectral evolution should be regarded as a qualitative indicator
of the crossover; the anomalous behavior itself is govened by the 
competition between $I_{N,c}$ and $I_{N,uc}$.

\begin{figure}[htbp]
\raggedleft
\includegraphics[width=8.5cm]{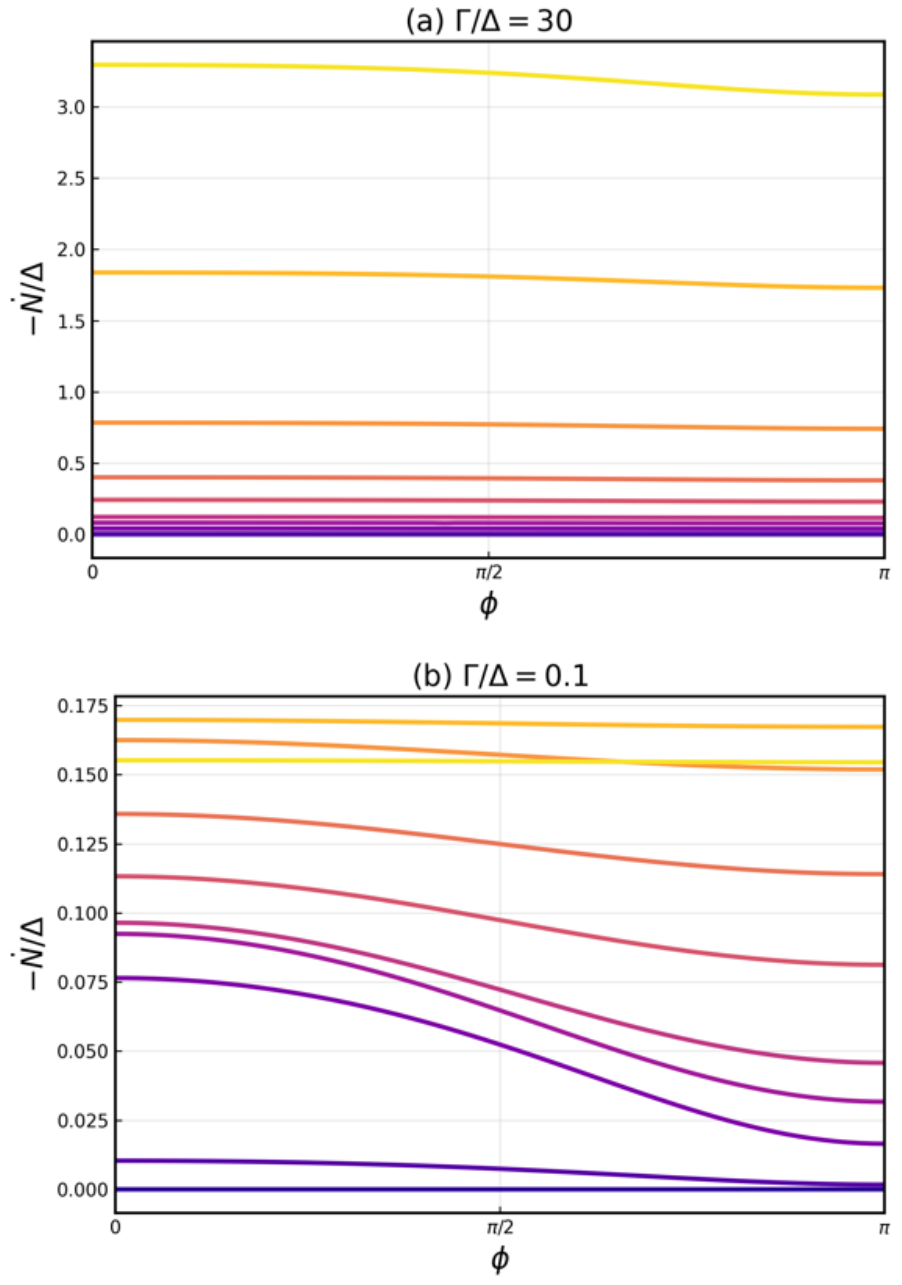}
\caption{\label{fig7}
Phase dependence of the loss current for spin-independent
loss at ${\cal T}=0.5$. (a) $\Gamma/\Delta=30$.
(b) $\Gamma/\Delta=0.1$. 
The color legend is same as that in Fig.~\ref{fig2}.}
\end{figure}

As  shown in Eq.~\eqref{eq:lcurrent},
the particle loss also produces a finite
 loss current.
In Fig.~\ref{fig7}, we show how the loss current varies as a function of 
the phase bias.
In the broad-resonance regime, the loss current increases substantially
with the loss rate (Fig.~\ref{fig7}~(a)).
In the narrow-resonance regime, its growth becomes much weaker at strong
loss (Fig.~\ref{fig7}~(b)), consistent with the onset of quantum-Zeno
suppression of the occupation of the lossy dot. 
Unlike the inter-reservoir particle and spin currents, the loss current
is even under phase reversal and is not required by symmetry
to vanish at $\phi=0$ or $\pi$. Its phase dependence is 
therefore considerably weaker than that of the supercurrents.

We now discuss a possible experimental
realization of the system proposed in this work.
In cold atoms, controllable fermionic superfluid junction have been realized
in mesoscopic transport setups~\cite{krinner2017}. 
In particular, digital mirror devices allow control over the conduction-channel geometry and can introduce particle loss in a spin-selective manner~\cite{PhysRevA.100.053605}. 
In condensed-matter systems,
a lossy junction may be realized in a three terminal geometry in which a mesoscopic sample is attached to two superconducting leads and one normal lead~\cite{PhysRevB.44.470,PhysRevB.62.1319,PhysRevA.106.053320}.
The single-particle loss can be mimicked when the chemical potential of the normal lead is much lower than that of the sample,
so that particle injection from the normal lead is negligible.
Spin-selective dissipation may also engineered using the ferromagnetic junctions~\cite{martin2001} or junctions with spin-orbit interactions~\cite{kohda2012}.
The Coulomb interaction in condensed-matter systems
may be incorporated, at the mean-field level, as a renormalization of the effective dot-level energy~\cite{martin2011}.

The $\pi$-junction properties
may be detected with
a DC SQUID(Superconducting Quantum Interference Device), which consists of a superconducting loop with two junctions and is available in both  cold atoms~\cite{PhysRevLett.111.205301,ryu2020quantum}
and condensed matter systems~\cite{PhysRevB.65.224513,persky2022studying}. 
In this setup,
the current-reversed properties
can be confirmed by replacing one of the junctions with
a $\pi$-junction.
This is because such a replacement is
equivalent to applying  half a flux quantum externally.
This method of confirming a $\pi$ junction has indeed 
been demonstrated in 
Ref.~\cite{PhysRevB.65.224513}.

We have discussed dissipation-induced emergent phenomena in simple mesoscopic systems.
Interesting future directions include studying
stability with respect to increasing the number of sites in the channel, 
correlation effects beyond mean field in a mesoscopic regime where connections to the Kondo effect~\cite{martin2011} and Tomonaga-Luttinger liquid are anticipated,
and extensions to bosonic superfluids, as realized in Josephson-junction array~\cite{PhysRevLett.116.235302}.

\section*{acknowledgment}
We thank T. Giamarchi and A.-M. Visuri for support in the early stage of
this work.
We are also grateful to T. Esslinger, P. Fabritius,
M.-Z. Huang, K. Kobayashi, J. Mohan, A. F. Morpurgo, M. Talebi, and
S. Wili for discussions.
SU is supported by JST PRESTO~(JPMJPR235) and
JSPS KAKENHI~(JP25K07191).

\appendix
\begin{widetext}
\section{Green's function}
The Keldysh components of real-time Green's functions between reservoir and dot are given by Eq.~\eqref{eq:KeldyshdLR} and Eq.~\eqref{eq:KeldyshLRd}. Similarly, real-time Green's functions at a quantum dot are expressed as
\begin{align}
    \hat{G}^R_d(\tau, \tau') &= -i \theta(\tau - \tau') \langle \{ \hat{d}(\tau), \hat{d}^\dagger(\tau') \} \rangle, \\
    \hat{G}^A_d(\tau, \tau') &= i \theta(\tau' - \tau) \langle \{ \hat{d}(\tau), \hat{d}^\dagger(\tau') \} \rangle, \\
    \hat{G}^K_d(\tau, \tau') &= -i \langle [ \hat{d}(\tau), \hat{d}^\dagger(\tau') ] \rangle.
\end{align}

We now obtain the current expression induced by the phase gradient between reservoirs.  By using the so-called Langreth rule~\cite{rammer2007},
we can obtain the following relations for 
Green's functions in frequency space~\cite{haug2008quantum}:
\begin{align}
    \hat{G}^K_{dL}(\omega) &= \hat{G}^R_d(\omega) \hat{t} \hat{g}^K_L(\omega) 
    + \hat{G}^K_d(\omega) \hat{t} \hat{g}^A_L(\omega), \\
    \hat{G}^K_{Ld}(\omega) &= \hat{g}^R_L(\omega) \hat{t}^\dagger \hat{G}^K_d(\omega) 
    + \hat{g}^K_L(\omega) \hat{t}^\dagger \hat{G}^A_d(\omega), \\
    \hat{G}^K_{dR}(\omega) &= \hat{G}^R_d(\omega) \hat{t}^\dagger \hat{g}^K_R(\omega) 
    + \hat{G}^K_d(\omega) \hat{t}^\dagger \hat{g}^A_R(\omega), \\
    \hat{G}^K_{Rd}(\omega) &= \hat{g}^R_R(\omega) \hat{t} \hat{G}^K_d(\omega) 
    + \hat{g}^K_R(\omega) \hat{t} \hat{G}^A_d(\omega).
\end{align}
Here, we introduce uncoupled retarded (advanced) Green's function in each reservoir $\hat{g}^{R}_{L(R)}(\omega)$ 
that does not incorporate
tunneling effects between reservoirs and quantum dot.
For superconducting reservoirs, it is given by~\cite{PhysRevB.54.7366}
\begin{align}
    \hat{g}^{R(A)}_{L(R)}(\omega) &\equiv
    \frac{1}{W}
    \begin{pmatrix}
        g^{R(A)}(\omega) & f^{R(A)}(\omega) \\
        f^{R(A)}(\omega) & g^{R(A)}(\omega)
    \end{pmatrix} \label{eq:uncoupled_g} \nonumber\\
    &=\frac{1}{W}
    \begin{pmatrix}
       -\frac{\omega \pm i0^+}{ \sqrt{\Delta^2 - (\omega \pm i0^+)^2}}  & \frac{\Delta}{ \sqrt{\Delta^2 - (\omega \pm i0^+)^2}}  \\
        \frac{\Delta}{ \sqrt{\Delta^2 - (\omega \pm i0^+)^2}}  & -\frac{\omega \pm i0^+}{ \sqrt{\Delta^2 - (\omega \pm i0^+)^2}} 
    \end{pmatrix}.
\end{align}
Since each reservoir is in equilibrium, the uncoupled Keldysh component is determined via the fluctuation-dissipation theorem~\cite{kamenev2011} as follows:
\begin{equation}
    \hat{g}^K_{L(R)}(\omega) = 
    \left[\hat{g}^R_{L(R)}(\omega) - \hat{g}^A_{L(R)}(\omega)\right] \left[1 - 2 n(\omega)\right],
\end{equation}
where $n(\omega) = \frac{1}{e^{\omega/T} + 1}$ is the Fermi distribution function.
We next look at uncoupled Green's function in the quantum dot. In the presence of the dissipation,  inverse Green's functions are determined as~\cite{PhysRevLett.130.200404}
\begin{align}
    \left[\hat{g}^{-1}_d(\omega)\right]^R &= 
    \begin{pmatrix}
        \omega - \epsilon + i \gamma_\uparrow & 0 \\
        0 & \omega + \epsilon + i \gamma_\downarrow
    \end{pmatrix} ,\nonumber\\
    \left[\hat{g}^{-1}_d(\omega)\right]^A &= 
    \begin{pmatrix}
        \omega - \epsilon - i \gamma_\uparrow & 0 \\
        0 & \omega + \epsilon - i \gamma_\downarrow
    \end{pmatrix} ,\nonumber\\
    \left[\hat{g}^{-1}_d(\omega)\right]^K &= 
    \begin{pmatrix}
        2 i \gamma_\uparrow & 0 \\
        0 & -2 i \gamma_\downarrow
    \end{pmatrix}.
\end{align}
Thus, uncoupled Green's functions are obtained as
\begin{align}
    \hat{g}^R_d(\omega) &= 
    \begin{pmatrix}
        \frac{1}{\omega - \epsilon + i \gamma_\uparrow} & 0 \\
        0 & \frac{1}{\omega + \epsilon + i \gamma_\downarrow}
    \end{pmatrix}, \\
    \hat{g}^A_d(\omega) &= 
    \begin{pmatrix}
        \frac{1}{\omega - \epsilon - i \gamma_\uparrow} & 0 \\
        0 & \frac{1}{\omega + \epsilon - i \gamma_\downarrow}
    \end{pmatrix} ,\\
    \hat{g}^K_d(\omega) &= 
    - \hat{g}^R_d(\omega) \left[\hat{g}^{-1}_d(\omega)\right]^K \hat{g}^A_d(\omega)=
    \begin{pmatrix}
        -\frac{2 i \gamma_\uparrow}{(\omega - \epsilon)^2 + \gamma_\uparrow^2} & 0 \\
        0 & \frac{2 i \gamma_\downarrow}{(\omega + \epsilon)^2 + \gamma_\downarrow^2}
    \end{pmatrix}.
\end{align}

\section{Dyson equation}
Since uncoupled Green's functions are determined, we turn to analyze the full Green's functions in the quantum dot $G_d$, obeying the following the Dyson equation~\cite{haug2008quantum}:
\begin{equation}
    \hat{G}_d(\omega) = \hat{g}_d(\omega) + \hat{g}_d(\omega) \hat{\Sigma}(\omega) \hat{G}_d(\omega)
    = \hat{g}_d(\omega) + \hat{G}_d(\omega) \hat{\Sigma}(\omega) \hat{g}_d(\omega),
\end{equation}
where $\hat{\Sigma}(\omega)$ is the self-energy.
We first look at the Dyson equation of the retarded and advanced components:
\begin{equation}
    \hat{G}_d^{R(A)}(\omega) = \hat{g}_d^{R(A)}(\omega) 
    + \hat{g}_d^{R(A)}(\omega) \hat{\Sigma}^{R(A)}(\omega) \hat{G}_d^{R(A)}(\omega),
\end{equation}
where
\begin{align}
    \hat{\Sigma}^{R(A)}(\omega) &= 
    \hat{t} \hat{g}_L^{R(A)}(\omega) \hat{t}^\dagger + \hat{t}^\dagger \hat{g}_R^{R(A)}(\omega) \hat{t}\nonumber\\
    &=\frac{\Gamma}{2}
    \begin{pmatrix}
        g_L^{R(A)}(\omega) + g_R^{R(A)}(\omega) 
        & -e^{-i\phi/2} f_L^{R(A)}(\omega) - e^{i\phi/2} f_R^{R(A)}(\omega) \\
        -e^{i\phi/2} f_L^{R(A)}(\omega) - e^{-i\phi/2} f_R^{R(A)}(\omega) 
        & g_L^{R(A)}(\omega) + g_R^{R(A)}(\omega)
    \end{pmatrix}.
\end{align}
The above Dyson equation can also be rewritten as follows:
\begin{equation}
    \left[\hat{G}_d^{R(A)}(\omega)\right]^{-1} = \left[\hat{g}_d^{R(A)}(\omega)\right]^{-1} - \hat{\Sigma}^{R(A)}(\omega).
\end{equation}
On the other hand, 
the Dyson equation of the Keldysh component is given by
\begin{align}
    \hat{G}_d^K(\omega) = 
    \left[1 + \hat{G}_d^R(\omega) \hat{\Sigma}^R(\omega)\right] \hat{g}_d^K(\omega) 
    \left[1 + \hat{\Sigma}^A(\omega) \hat{G}_d^A(\omega)\right]+ \hat{G}_d^R(\omega) \hat{\Sigma}^K(\omega) \hat{G}_d^A(\omega).
\end{align}
Here
\begin{align}
\hat{\Sigma}^K(\omega)&=  \hat{t} \hat{g}_L^K(\omega) \hat{t}^{\dagger}+\hat{t}^{\dagger} \hat{g}_R^K(\omega) \hat{t} \nonumber\\ 
&={\left[\hat{\Sigma}^R(\omega)-\hat{\Sigma}^A(\omega)\right][1-2 n(\omega)] } \nonumber\\ 
&= {\left[\left[\hat{g}_d^R(\omega)\right]^{-1}-\left[\hat{g}_d^A(\omega)\right]^{-1}\right][1-2 n(\omega)]-\left[\left[\hat{G}_d^R(\omega)\right]^{-1}-\left[\hat{G}_d^A(\omega)\right]^{-1}\right][1-2 n(\omega)] } .
\end{align}
By using above and the Dyson equation for $G_d^{R(A)}$, 
$\hat{G}^K_d$ is rewritten as
\begin{align}
\hat{G}_d^K(\omega)
=&\hat{G}_d^R(\omega)\left[\hat{g}_d^R(\omega)\right]^{-1} \hat{g}_d^K(\omega)\left[\hat{g}_d^A(\omega)\right]^{-1} \hat{G}_d^A(\omega)+\hat{G}_d^R(\omega) \hat{\Sigma}^K(\omega) \hat{G}_d^A(\omega) \nonumber\\
=& -\hat{G}_d^R(\omega)\left[\hat{g}_d^{-1}(\omega)\right]^K 
\hat{G}_d^A(\omega)+\hat{G}_d^R(\omega)\left[\hat{\Sigma}^R(\omega)-\hat{\Sigma}^A(\omega)\right] \hat{G}_d^A(\omega)[1-2 n(\omega)] \nonumber\\ 
=& -\hat{G}_d^R(\omega)\left[\hat{g}_d^{-1}(\omega)\right]^K \hat{G}_d^A(\omega)+\left[\hat{G}_d^R(\omega)-\hat{G}_d^A(\omega)\right][1-2 n(\omega)]\nonumber\\
&+\hat{G}_d^R(\omega)\left[\left[\hat{g}_d^R(\omega)\right]^{-1}-\left[\hat{g}_d^A(\omega)\right]^{-1}\right]\hat{G}_d^A(\omega)[1-2 n(\omega)].
\end{align}

\section{Current expressions}
Here, we derive the current expressions utilized
for numerics in this work.

By means of the Dyson equations obtained above, 
the supercurrent $I_N$ is expressed as
\begin{align}
I_N=&-\int \frac{d\omega}{8\pi} \mathrm{Tr} \bigg[- \hat{\sigma}_z \hat{t}^\dagger \big\{ \hat{G}_d^R \hat{t}\hat{g}_L^K + \hat{G}_d^K \hat{t}\hat{g}_L^A \big\}+ \hat{\sigma}_z \hat{t} \big\{ \hat{g}_L^R \hat{t}^\dagger \hat{G}_d^K + \hat{g}_L^K \hat{t}^\dagger \hat{G}_d^A \big\}\nonumber\\
&+\hat{\sigma}_z \hat{t} \big\{ \hat{G}_d^R \hat{t}^\dagger\hat{g}_R^K + \hat{G}_d^K \hat{t}^\dagger\hat{g}_R^A \big\} \hat{t}- \hat{\sigma}_z \hat{t}^\dagger \big\{ \hat{g}_R^R \hat{t} \hat{G}_d^K + \hat{g}_R^K \hat{t} \hat{G}_d^A \big\} 
\bigg] \nonumber\\
=&I_{N,c}+I_{N,uc},
\end{align}
where
\begin{align}
I_{N,c}=&-\int \frac{d\omega}{8\pi} \mathrm{Tr} \bigg[- \hat{\sigma}_z \hat{t}^\dagger \big[ \hat{G}_d^R \hat{t}\hat{g}_L^R - \hat{G}_d^A \hat{t}\hat{g}_L^A \big] 
  + \hat{\sigma}_z \hat{t} \big[ \hat{g}_L^R \hat{t}^\dagger \hat{G}_d^R - \hat{g}_L^A \hat{t}^\dagger \hat{G}_d^A \big]\nonumber\\
  &+\hat{\sigma}_z \hat{t}^\dagger \big[ \hat{G}_d^R \hat{t}^\dagger \hat{g}_R^R - \hat{G}_d^A \hat{t}^\dagger \hat{g}_R^A \big] 
  - \hat{\sigma}_z \hat{t}^\dagger \big[ \hat{g}_R^R \hat{t} \hat{G}_d^R - \hat{g}_R^A \hat{t}\hat{G}_d^A \big]
\bigg] \big[ 1 - 2n(\omega) \big],\label{eq:particle-currentc1}\\
I_{N,uc}=&-\int \frac{d\omega}{8\pi} \mathrm{Tr} \bigg[\hat{\sigma}_z \hat{t}^\dagger \hat{G}_d^R(\omega) [\hat{g}_d^{-1}(\omega)]^K \hat{G}_d^A (\omega) \hat{t} \hat{g}_L^A- \hat{\sigma}_z \hat{t}\hat{g}_L^R\hat{t}^\dagger\hat{G}_d^R(\omega)[\hat{g}_d^{-1}(\omega)]^K \hat{G}_d^A (\omega)\nonumber\\
&-\hat{\sigma}_z \hat{t} \hat{G}_d^R(\omega)  [\hat{g}_d^{-1}(\omega)]^K \hat{G}_d^A (\omega) \hat{t}^\dagger \hat{g}_R^A +\hat{\sigma}_z \hat{t}^\dagger \hat{g}_R^R\hat{t} \hat{G}_d^R(\omega) [\hat{g}_d^{-1}(\omega)]^K \hat{G}_d^A (\omega)\bigg] \nonumber \\
&-\mathrm{Tr} \bigg[-\hat{\sigma}_z \hat{t}^\dagger \hat{G}_d^R(\omega)([\hat{g}_d^R (\omega)]^{-1}-[\hat{g}_d^A (\omega)]^{-1})\hat{G}_d^A (\omega) \hat{t} \hat{g}_L^A+\hat{\sigma}_z \hat{t}\hat{g}_L^R\hat{t}^\dagger\hat{G}_d^R(\omega)([\hat{g}_d^R (\omega)]^{-1}-[\hat{g}_d^A (\omega)]^{-1})\hat{G}_d^A (\omega)\nonumber\\
&+\hat{\sigma}_z \hat{t} \hat{G}_d^R(\omega)([\hat{g}_d^R (\omega)]^{-1}-[\hat{g}_d^A (\omega)]^{-1})\hat{G}_d^A (\omega) \hat{t}^\dagger \hat{g}_R^A-\hat{\sigma}_z \hat{t}^\dagger \hat{g}_R^R\hat{t}\hat{G}_d^R(\omega)([\hat{g}_d^R (\omega)]^{-1}-[\hat{g}_d^A (\omega)]^{-1})\hat{G}_d^A (\omega)\bigg]\big[ 1 - 2n(\omega) \big].\label{eq:particle-currentc2}
\end{align}

We point out that in the limit of $\gamma\to0$ and
$\Gamma/\Delta\gg1$, the expression above
reduces to the Andreev current $I_{\text{ABS}}$,
that is,
\begin{align}
I_N&\to I_{\text{ABS}}\nonumber\\
&=\frac{{\cal T}\Delta^2\sin\phi}{2E_{\text{ABS}}}
\tanh\left(\frac{E_{\text{ABS}}}{2T}\right),
\end{align}
where
\begin{align}
E_{\text{ABS}}=\Delta\sqrt{1-{\cal T}\sin^2(\phi/2)}
\end{align}
is the so-called Andreev bound-state energy.

\begin{figure*}[htbp]
    \centering
    \includegraphics[width=15cm]{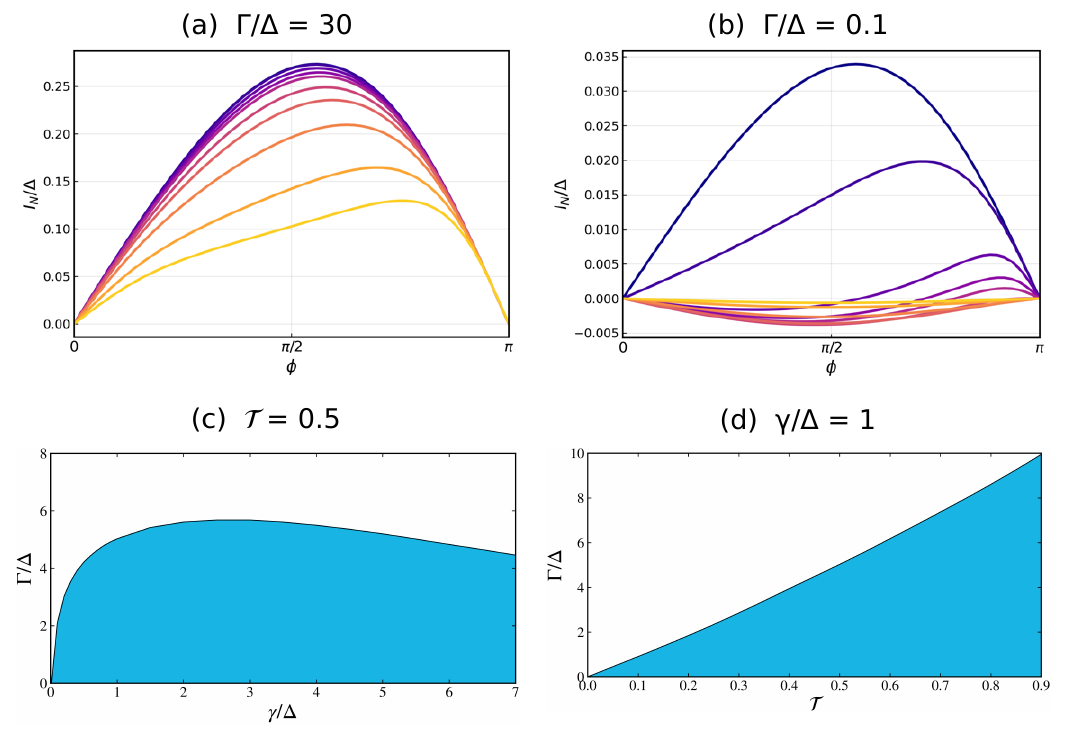}
    \caption{\label{fig8}
    Particle supercurrent generated by spin-selective loss, $\gamma_{\uparrow}=\gamma$ and $\gamma_{\downarrow}=0$.
    (a,b) $I_N(\phi)$ at ${\cal T}=0.5$ for $\Gamma/\Delta=30$
    and 0.1, respectively; the curves use the $\gamma/\Delta$ values listed in
    Fig.~\ref{fig5}.
    (c,d) Phase diagrams at fixed ${\cal T}=0.5$ and fixed
    $\gamma/\Delta=1$, respectively. Blue regions occur the anomalous behaviors in the current-phase relation.
    }
\end{figure*} 

The spin supercurrent between reservoirs $I_S$ is expressed as
\begin{align}
I_S=&I_{S,c}+I_{S,uc},
\end{align}
where
\begin{align}
I_{S,c}=&-\int \frac{d\omega}{8\pi} \mathrm{Tr} \bigg[-\hat{t}^{\dagger}\left[\hat{G}_d^R \hat{t} \hat{g}_L^R-\hat{G}_d^A \hat{t} \hat{g}_L^A\right]+\hat{t}\left[\hat{g}_L^R \hat{t}^{\dagger} \hat{G}_d^R-\hat{g}_L^A \hat{t}^{\dagger} \hat{G}_d^A\right]\nonumber\\
&+\hat{t}\left[\hat{G}_d^R \hat{t}^{\dagger} \hat{g}_R^R-\hat{G}_d^A \hat{t}^{\dagger} \hat{g}_R^A\right]-\hat{t}^{\dagger}\left[\hat{g}_R^R \hat{t} \hat{G}_d^R-\hat{g}_R^A \hat{t} \hat{G}_d^A\right]\bigg][1-2 n(\omega)],\\
I_{S,uc}=&-\int \frac{d\omega}{8\pi} \mathrm{Tr} \bigg[\hat{t}^{\dagger} \hat{G}_d^R(\omega)\left[\hat{g}_d^{-1}(\omega)\right]^K \hat{G}_d^A(\omega) \hat{t} \hat{g}_L^A-\hat{t} \hat{g}_L^R \hat{t}^{\dagger} \hat{G}_d^R(\omega)\left[\hat{g}_d^{-1}(\omega)\right]^K \hat{G}_d^A(\omega)\nonumber\\
&-\hat{t} \hat{G}_d^R(\omega)\left[\hat{g}_d^{-1}(\omega)\right]^K \hat{G}_d^A(\omega) \hat{t}^{\dagger} \hat{g}_R^A+\hat{t}^{\dagger} \hat{g}_R^R \hat{t} \hat{G}_d^R(\omega)\left[\hat{g}_d^{-1}(\omega)\right]^K \hat{G}_d^A(\omega)\bigg] \nonumber\\
&-\operatorname{Tr}\bigg[-\hat{t}^{\dagger} \hat{G}_d^R(\omega)\left(\left[\hat{g}_d^R(\omega)\right]^{-1}-\left[\hat{g}_d^A(\omega)\right]^{-1}\right) \hat{G}_d^A(\omega) \hat{t}\hat{g}_L^A+\hat{t}\hat{g}_L^R \hat{t}^{\dagger} \hat{G}_d^R(\omega)\left(\left[\hat{g}_d^R(\omega)\right]^{-1}-\left[\hat{g}_d^A(\omega)\right]^{-1}\right) \hat{G}_d^A(\omega)\nonumber\\
&+\hat{t} \hat{G}_d^R(\omega)\left(\left[\hat{g}_d^R(\omega)\right]^{-1}-\left[\hat{g}_d^A(\omega)\right]^{-1}\right) \hat{G}_d^A(\omega) \hat{t}^{\dagger} \hat{g}_R^A-\hat{t}^{\dagger} \hat{g}_R^R \hat{t} \hat{G}_d^R(\omega)\left(\left[\hat{g}_d^R(\omega)\right]^{-1}-\left[\hat{g}_d^A(\omega)\right]^{-1}\right) \hat{G}_d^A(\omega)\bigg][1-2 n(\omega)].
\label{eq:spin-currentc}
\end{align}
The direct calculation shows
\begin{align}
&\operatorname{Tr}\left[-\hat{t}^{\dagger} \hat{G}_d^{R(A)} \hat{t}\hat{g}^{R(A)}+\hat{t}\hat{g}^{R(A)} \hat{t}^{\dagger} \hat{G}_d^{R(A)}\right]=0,\\
&\operatorname{Tr}\left[-\hat{t} \hat{G}_d^{R(A)} \hat{t}^{\dagger}\hat{g}^{R(A)}+\hat{t}^{\dagger} \hat{g}^{R(A)} \hat{t} \hat{G}_d^{R(A)}\right]=0,
\end{align}
which implies that $I_{S,c}$ vanishes. This result shows that the spin supercurrent is fully induced by the term associated with the dissipation effect. Thus, the spin supercurrent is expressed as
\begin{align}
I_S=I_{S,uc}.
\end{align}

By  Fourier transforming Eq.~\eqref{eq:lcurrent2}
the loss current is expressed as
\beq
-\dot{N}=i\int\frac{d\omega}{2\pi}\text{Tr}\Big[
\hat{\gamma}\{\hat{G}^R_d(\omega)-\hat{G}^A_d(\omega)\}
-\hat{\gamma}\hat{\sigma}_z\hat{G}^K_d(\omega)
\Big].
\label{eq:lcurrentc}
\eeq
In the main text, we numerically evaluate Eqs.~\eqref{eq:particle-currentc1},
~\eqref{eq:particle-currentc2},~\eqref{eq:spin-currentc}, and ~\eqref{eq:lcurrentc}.
\end{widetext}

\section{Supercurrent in the presence of spin-selective loss}
Here we demonstrate the particle supercurrent behaviors under spin-selective loss, with $\gamma_{\uparrow}=\gamma$ and $\gamma_{\downarrow}=0$.
\par
Figures~\ref{fig8}~(a) and (b) show the current-phase relation
at several loss strengths in the broad- and narrow-resonance regimes, respectively.
As in the case of the spin-independent case, the supercurrent decreases
monotonically with $\gamma$ in the broad-resonance regime,
whereas anomalous current-phase relations appear for $\Gamma/\Delta\lesssim1$.
\par
In addition, phase diagrams of the particle supercurrent with fixing transparency and
dissipation strength are respectively shown in Fig.~\ref{fig8}~(c) and (d).
As in the spin-independent case, the anomalous regime initially
expands with $\gamma$ but shrinks again for $\gamma/\Delta\gtrsim2$.
By contrast, it grows monotonically with transmission.

\bibliographystyle{apsrev4-1}
\bibliography{reference}

\end{document}